\documentclass[aps,preprint,showpacs,superscriptaddress,groupedaddress]{revtex4}  
\usepackage[dvips]{graphicx}
\usepackage{dcolumn}   
\usepackage{bm}        
\usepackage{amssymb}   
\usepackage[utf8]{inputenc}
\usepackage{mathrsfs}
\usepackage{amsmath}
\usepackage[english]{babel}
\usepackage{hyperref}
\usepackage{color}

\begin{document}
\title{Double charged heavy constituents of dark atoms and superheavy nuclear objects}
\author{Vakhid A. Gani}
\email{vagani@mephi.ru}
\affiliation{National Research Nuclear University MEPhI (Moscow Engineering Physics Institute), 115409 Moscow, Russia}
\affiliation{National Research Center Kurchatov Institute, Institute for Theoretical and Experimental Physics, 117218 Moscow, Russia}
\author{Maxim Yu. Khlopov}
\email{khlopov@apc.univ-paris7.fr}
\affiliation{Institute of Physics,
Southern Federal University\\
Stachki 194,
Rostov on Don 344090, Russia}
\author{Dmitry N. Voskresensky}
\email{d.voskresen@gmail.com}
\affiliation{National Research Nuclear University MEPhI (Moscow Engineering Physics Institute), 115409 Moscow, Russia}
\affiliation{Joint Institute for Nuclear Research, RU-141980 Dubna, Moscow region, Russia}
\vskip 1cm
\begin{abstract} We consider the model of composite dark matter assuming stable particles of charge $-2$ bound with primordial helium nuclei by the Coulomb force in $O$He atoms. We study capture of such dark atoms in matter and propose the possibility of the existence of  stable $O$-enriched superheavy nuclei and $O$-nuclearites, in  which heavy $O$-dark matter fermions  are bound by electromagnetic forces with ordinary  nuclear matter. $O$He atoms accumulation in stars and its possible effect on stellar evolution is also considered, extending the set of indirect probes for composite dark matter.
\end{abstract}
\maketitle


\section{Introduction}
\label{sec:sec1}

There is overwhelming evidence for the presence of a dark matter (DM) in the Universe \cite{Ade:2015xua} and together with most popular, but still elusive weakly interacting massive particle (WIMP) \cite{Arcadi.2018}, there exist numerous theoretical models including axions, sterile neutrinos, primordial black holes \cite{Belotsky.2014,Carr.2016,Carr.PRD.2010}, strongly interacting massive particles and superweakly interacting particles (see Refs.~\cite{DMRev,Aprile:2009zzd,Feng:2010gw} for review and references). Even  electromagnetically interacting massive particle (EIMP) candidates are possibly hidden in neutral atomlike states. Dark $O$He atoms, in which hypothetical $-2$ charged particles are bound with primordial helium nuclei, occupy a special place on this list. Such models involve only one free parameter of new physics --- the mass of $-2$ charged EIMPs --- so many features of this type of dark matter can be described by the known nuclear and atomic physics.

In 2005, Glashow \cite{Glashow:2005jy} proposed a kind of EIMP model, according to which stable teraquarks $U$ (of mass of the order of tera-electron-volts and of electric charge $+2/3$) form a $UUU$ baryon bound with tera-electrons $E$ of charge $-1$ in the neutral $(UUUEE)$ atom. However, the primordial He formed in the big bang nucleosynthesis captures all the free $E$ in positively charged (He$E$)$^+$ ions, preventing a required suppression of the positively charged particles that can bind with electrons in atoms of anomalous hydrogen. In general, stable single charged EIMPs form anomalous hydrogen either directly binding with ordinary electrons ($+1$ charged EIMPs), or indirectly ($-1$ charged EIMPs) forming first $+1$ charge ion with primordial helium and then anomalous hydrogen with ordinary electrons \cite{BKSR1}. Therefore, anomalous hydrogen overproduction excludes any significant amount of stable single charged EIMPs.

Nevertheless, there are several models that predict stable double charged particles without stable single charged particles. In particular, the hypothesis of the heavy stable quark of the fourth family may provide a solution, if an excess of $\bar{U}$ antiquarks with charge $(-2/3)$ is generated in the early Universe. Excessive $\bar{U}$ antiquarks then form $\bar{U}\bar{U}\bar{U}$ antibaryons with the electric charge $-2$, which are captured by He forming $O^{--}$He$^{++}$ ($O$He) atoms \cite{Khlopov:2005ew} right after the appearance of the He nuclei in the big bang nucleosynthesis. This hypothesis has found implementations in the model of almost commutative geometry as well as in models of walking technicolor and has been extensively discussed in the literature; see Refs.~\cite{Khlopov:2010ik,Khlopov:2011me,gKhlopov,khlopov.proc.2014,khlopov.proc.2015,khlopov.ijmpd.2015,Wallemacq:2015cjr} and references therein. The model is particularly predictive since the only parameter that one needs to know is the mass of the $O$-particle. The model can explain the observed excess of the positronium annihilation line in the galactic bulge and excessive fraction of high-energy cosmic-ray positrons, if the mass of this particle does not exceed 1.3 TeV, challenging the direct test of this explanation in searches for stable double charged particles at the LHC \cite{probes}.

Charge conservation implies the existence of $+2$ charged particle $O^{++}$ together with $O^{--}$. To avoid overproduction of anomalous isotopes by $O^{++}$, $O$He-dominated dark matter should be asymmetric with strongly suppressed $+2$ charged particles. In the walking technicolor model \cite{KK,KK1}, due to sphaleron transitions, such excess is related to the baryon excess, giving the observed dark matter/baryon matter density ratio for a reasonable choice of parameters.

In the early Universe when temperature fell below 1 keV, the rate of expansion started to exceed the rate of energy and momentum transfer from plasma to $O$He gas (see, e.g., Ref.~\cite{gKhlopov} for review and references). As a result, $O$He decoupled from plasma and radiation and played the role of dark matter on the matter-dominated stage. Before decoupling from plasma and radiation, $O$He density fluctuations convert in sound waves. It leads to the suppression of small-scale fluctuations. Thereby $O$He dark matter was called warmer than cold dark matter for an $O$He mass about 1 TeV, typical for cold dark matter particles \cite{khlopov.proc.2014}. The averaged baryonic density in the course of structure formation and in galaxies is sufficiently low making baryonic matter at large scales transparent for $O$He. So, for a galaxy with mass $M = 10^{10}M_\odot$ and radius $R = 10^{23}$ cm, $n\sigma R = 8 \cdot 10^{-5} \ll 1$, where $n = M/4\pi R^3$ and $\sigma = 2 \cdot 10^{-25}$ cm$^2$ is the geometrical cross section for $O$He collisions. For that reason, in the period of formation of the first objects, $O$He does not follow the condensation of baryonic matter, so the $O$He model avoids constraints from the cosmic microwave background \cite{CMB} and formation of the first stars \cite{FS}. In galaxies and galaxy clusters, $O$He behaves like collisionless gas avoiding constraints from Bullet Cluster observations \cite{BC}. Only dense matter objects like stars or planets are opaque for it. The protostellar cloud with the solar mass becomes opaque for $O$He when it contracts within $8 \cdot 10^{15}$ cm. Correspondingly, the protoplanet cloud of the mass of the Earth becomes opaque when it contracts to $10^{13}$ cm. 

Because of the nuclear interaction cross section of elastic collisions with terrestrial matter, $O$He is slowed down to thermal velocity in the matter of underground detectors. It leads to negligible nuclear recoil in $O$He collisions with nuclei in direct-detection experiments. Positive results of DAMA/NaI and DAMA/LIBRA and negative results of other groups are explained in the $O$He model by annual modulation of the rate of low-energy binding of $O$He with intermediate mass nuclei \cite{Khlopov:2010ik,Khlopov:2011me,gKhlopov}. Open problems of this explanation related with the existence and role of the dipole potential barrier in $O$He-nucleus interaction are discussed in Refs.~\cite{khlopov.proc.2014,khlopov.proc.2015,khlopov.ijmpd.2015,Wallemacq:2015cjr}.

On the other hand, various hypotheses of the existence  of superheavy nuclei with the atomic numbers essentially higher than that of ordinary atomic nuclei have been explored. In 1971, Migdal suggested the possibility of superdense nuclei glued by a pion condensate \cite{Migdal:1971cu,Migdal:1974yx,Migdal:1978az,Migdal:1990vm}. Lee and Wick conjectured $\sigma$-condensate superheavy nuclei \cite{Lee:1974ma,Lee:1974kn}. Bodmer proposed collapsed quark nuclei \cite{Bodmer:1971we}. Reference \cite{Migdal:1977rn} demonstrated that the interior of a nucleus with a charge $Z\gg 1/e^3$, $e$ is the charge of the electron, $\hbar=c=1$, is electrically neutral and Refs.~\cite{Voskresensky:1977mz,Voskresensky:1978uf,Migdal:1990vm} suggested the possibility of existence of nuclei stars of the atomic number $(10^2-10^3)\le A\le 10^{57}$, the electric charge of which is compensated by the negatively charged pion condensate and the electrons. References \cite{Migdal:1990vm,Voskresensky1977,Kolomeitsev:2002gd} argued that if there existed negatively charged light bosons of mass less than $(30-32)$ MeV there would exist exotic objects, nuclei stars, of arbitrary size (until the effects of gravity can be neglected) with density typical for normal atomic nuclei, bound by strong and electromagnetic interactions. Witten \cite{Witten:1984rs} suggested the possible existence of quark nuggets, constructed from up, down, and strange quarks, with the atomic number between $(3\cdot 10^2-10^3)\le A\le 10^{57}$, see Ref.~\cite{Alcock:1986hz}, as candidates for the DM in the Universe. De Rujula and Glashow \cite{DeRujula:1984axn} called these stable drops ``nuclearites'' and discussed conditions for their feasible detection in terrestrial conditions. They have also discussed charged massive particles (CHAMPs) \cite{DeRujula:1989fe}. They argued that negative CHAMPs may bind to protons in superheavy isotopes. Superheavy nuclei and nuclearites may exist in the Galaxy as debris from the big bang, supernovae explosions, star collisions, and other astrophysical catastrophes. Numerous subsequent works focused on the consideration of the strange stars as a new family of compact stars. Besides that, exotic matter like the pion condensates and the quark matter in various phases may exist in the interiors of some neutron stars \cite{Weber,Ivanov:2005be,Boeckel:2010hm,Dondi:2016yjl}. The other side of the problem is the possible influence of dark matter captured by stars on the stellar structure and evolution. In particular, it can lead to observable effects in neutron stars \cite{Kouvaris.PRD.2014}.

Below, we assume that the DM may consist of $O$-particles bound in $O$He atoms. Colliding with the ordinary atomic nuclei, $O$He atoms may undergo fusion reactions with the formation of superheavy $O$-nuclei. However, the simplest from the viewpoint of new physics and principally being the subject of the complete quantum mechanical treatment of $O$He interaction with matter, such a description still remains an open question of the $O$He model. Putting aside this uncertainty, we suggest the idea of the possibility of the existence of $O$-nuclearites, constructed of self-bound nuclear matter at the density typical for the nuclear saturation, in which the positive electric charge of protons is compensated by negatively charged $O^{--}$. Such nuclearites might be formed in $O$He interaction with nuclei, and we study their effect in astrophysical conditions.

The paper is organized as follows. In Sec.~\ref{sec:sec2}, we formulate the idea of the existence of $O$-nuclearites. In Sec.~\ref{sec:sec3} we take into account the effects of gravitation. Then, Sec.~\ref{sec:sec4} presents some estimates for the $O$-nuclearite accumulation during star evolutions. Finally, Sec.~\ref{sec:sec5} contains some concluding remarks.

\section{Self-bound $O$-nuclearites}
\label{sec:sec2}

Consider an ordinary atomic nucleus of atomic number $A$. Assume that we deal with a rather heavy nucleus of isospin-symmetric composition (the number of neutrons $N_\mathrm{n}$ is equal to the number of protons $N_\mathrm{p}$, $A=2N_\mathrm{p}$). Then the proton and neutron densities are $n_\mathrm{p} = n_\mathrm{n} = n_\mathrm{p}^0 \theta(r-R)$, except a narrow nuclear diffuseness layer $\delta R\sim 0.5$ fm near the surface, $2n_\mathrm{p}^0 = n_0 = 0.16$ fm$^{-3}$ is the normal nuclear density. Assume that there is a distribution of heavy $O$-particles inside the nucleus with a density $n_O (r)$. This approach differs from early studies of bound systems of stable heavy negatively charged particles with nuclei \cite{Cahn,Pospelov,Kohri}.

The energy of such a constructed $O$-nuclearite is
\begin{equation}\label{E}
{\cal{E}}=-16 {\rm{
 MeV}} \cdot A  - \int d^3 r (n_\mathrm{p} -2n_O)V -\int d^3 r\frac{(\nabla{V})^2}{8\pi e^2} + {\cal{E}}_{\rm kin}^{O}\,.
\end{equation}
Here, the first term is the volume energy of the atomic nucleus, the next two terms describe the electromagnetic energy, and
\begin{equation}\label{E_kin}
{\cal{E}}_{\rm kin}^O = \displaystyle\int d^3 r \int\limits_0^{p_{{\rm F},O}^{}} \frac{p^2 dp}{\pi^2}\frac{p^2}{2m_O^{}}
\end{equation}
is the kinetic energy of the $O$-fermions of the mass $m_O$; $V = -e\phi$ is the potential well for the electron in the field of the positive charge ($e>0$, $\phi>0$), and on the other hand, it is the potential well also for the protons in the field of the negative charge of $O$-particles, $n_O^{}=p_{{\rm F},O}^3/(3\pi^2)$, $n_{\rm p} = n_\mathrm{n} = p_{{\rm F},p}^3/(3\pi^2)$, $p_{{\rm F},p}^{}\simeq \sqrt{2m_\mathrm{N}^{}|V|}$; see Ref.~\cite{Migdal:1977rn}. The contribution of ${\cal{E}}_{\rm kin}^{O}$ is tiny, provided $m_O\gg m_{\rm N}$, where $m_{\rm N}$ is the nucleon mass (following Ref.~\cite{Khlopov:2005ew}, in our estimations, we assume $m_O\simeq $TeV), and can be neglected along with the nuclear surface term arising due to a redistribution of the charge in a narrow diffuseness layer.

The charge distribution can be found from the Poisson equation
\begin{eqnarray}\label{Poisson}
\Delta V=4\pi e^2 (n_p - 2n_O)
\end{eqnarray}
obtained from the minimization of the energy. Multiplying Eq.~\eqref{Poisson} by $V$ and integrating it out, we find that the Coulomb part of the total energy $\int d^3 r\displaystyle\frac{(\nabla{V})^2}{8\pi e^2}$ is always non-negative. Thus, the most energetically favorable $O$-particle distribution inside the nucleus should fully compensate the Coulomb field, following the proton distibution. Thereby, $O$-particles, if their number were $N_O\geq A/4$, would be redistributed to minimize the energy, and finally the density of $O$ inside the atomic nucleus becomes $n_O = n_\mathrm{p}/2 = (n_\mathrm{p}^0/2)\:\theta(r-R)$ for the $O$-nuclearite, which corresponds to $V = const$ for $r<R$. Excessive $O$-particles are pushed out. Thus, the constructed $O$-nuclearite has the energy ${\cal{E}}\simeq -16 {\rm{MeV}} \cdot A<0$, and thereby, for arbitrary $A$, it proves to be absolutely stable (if $O$ is considered as a stable particle), untill gravity is yet unimportant. The assumption $n_\mathrm{p} = n_\mathrm{n} = n_\mathrm{p}^0\:\theta(r-R)$ made above is actually not necessary; the key point here is that it is profitable to have $n_O (r) = n_\mathrm{p}(r)/2$, if there is a sufficient amount of $O$-particles.

Note that the value ${\cal{E}}_{\rm kin}^O<16 $MeV$\cdot A$, and thereby, the matter of the nuclearite is self-bound, provided $m_O >2.3 m_N$. Also note that we considered nuclearites of which the electric charge is compensated by $O^{--}=\bar{U}\bar{U}\bar{U}$. On equal footing, we could consider antinuclearites made of antiprotons and antineutrons at typical density $n\sim n_0$ with the electric charge compensated  by $O^{++}=UUU$.

\section{Self-gravitating $O$-nuclearites and black holes}
\label{sec:sec3}

With increase of $A$, the gravity comes into play. The density profile can be found from the solution of the Tolman-Oppenheimer-Volkoff equation. However, even not solving this  equation, we are able to roughly estimate the typical size of the gravitationally stable $O$-nuclearite, similarly to the estimation valid for neutron stars. We assume that ${\cal{E}}_{\rm kin}^{\rm nucl}\ge |{\cal{E}}_{\rm pot}^{\rm nucl}|$, for typical densities under consideration. Then the internal pressure is determined by the Fermi gas of the nucleons. The corresponding energy term is $\sim {\cal{E}}_{\rm kin}^{\rm nucl}\sim (p_{{\rm F}, \rm nucl}^2/m_\mathrm{N}) A$, $p_{{\rm F},\rm nucl}\sim A^{1/3}/R$. The gravitational energy is ${\cal{E}}_{\rm grav}\sim -GM^2/R$, $M\simeq A m_O$. In a gravitationally stable object, the internal (nucleon) pressure is compensated by the gravitational one. Thus, we estimate
\begin{eqnarray}\label{R}
R\sim 1/(GM^{1/3}m_\mathrm{N} m_O^{5/3})\sim 10 {\rm{km}} (M_{\odot}/M)^{1/3}(m_\mathrm{N}/m_O)^{5/3}\,.
\end{eqnarray}
For an individual self-gravitating $O$-nuclearite to remain in a self-bounded state the nucleon density should be $n < (2-2.5) n_0$ since for realistic equations of state at such baryon densities the energy of the isosymmetric nuclear matter (at the switched-off Coulomb term) remains negative; see Fig.~1 in Ref.~\cite{Klahn:2006ir}. Assuming for a rough estimate that the internal pressure is of the order of that for the ideal Fermi gas of nucleons, from Eq.~\eqref{R}, we find that, in order for an individual $O$-nuclearite to have central density $n\sim (2-2.5) n_0$, its  mass should be $M\sim 3\cdot 10^{-8}M_{\odot}$ and the radius $R\sim 30$ m (for $m_O\simeq 10^3 m_\mathrm{N}$ that we use).

With the increase of the $O$-nuclearite mass, the central density continues to increase. From the condition $R>R_\mathrm{G} = 2G M\sim 4 (M/M_{\odot}){\rm{km}}$, we may estimate the maximum available mass of the $O$-nuclearite to not become a black hole. For $m_O\sim 10^3 m_\mathrm{N}$ equating $R$ and $R_\mathrm{G}$, we estimate $M_{\rm max}\sim 0.3 \cdot 10^{-3}M_{\odot}$, $R_{\rm min}\sim 10^{-3}$ km that corresponds to the central density $n_{\rm max}\sim 10^5 n_0$. For $M>M_{\rm max}$, the $O$-nuclearite would become a black hole.

The masses of neutron stars are assumed to vary in the range $0.7 M_{\odot}\le M\le (2-3)M_{\odot}$. Thus, passing through a flux composed of $O$He atoms, $O$-nuclei and  $O$-nuclearites, a neutron star of the mass $M\ge M_{\odot}$ during its evolution may accumulate  at most $\sim 10^3$ of the most heavy $O$-nuclearites (of total mass $\sim M_\odot$ as we have estimated above) before it converts  into the black hole.

Note that the local density of a nonluminous mass in the galaxies is $\rho_\mathrm{DM}^{} \simeq (3-7)\cdot 10^{-25}$g$/$cm$^3$ \cite{Chin:1979yb}. We further assume that $\rho_\mathrm{DM}^{} \simeq \rho_{O{\rm He}}^{}$  and that interactions of $O$He with ordinary matter are dominantly elastic. However, if the $O$-particle enters inside an ordinary nucleus, it is energetically profitable for it to remain there, making the nucleus superheavy. Thus, absorption of $O$ and $\alpha$ particles in inelastic collisions of cosmic $O$He with nuclei yields with some probability new $\{OA\}_{N_\mathrm{n}}^{N_\mathrm{p}}$ nuclei. Such events should be very rare at least since no one $O$-nucleus has been observed yet, and a mechanism for $O$-nuclearite formation should be rather peculiar.

\section{Accumulation of $O$-nuclearites during the star evolution}
\label{sec:sec4}

To be specific, consider the accretion of $O$He flux onto a neutron star. Masses of neutron stars with central densities $n_{\rm cen}\le (2-2.5) n_0$ vary typically between $(0.7-1.8)M_{\odot}$, and the specific values depend on the choice of the  equation of state of the neutron-star matter; see Fig.~2 in Ref.~\cite{Klahn:2006ir}. Self-bound $O$-nuclearites might be formed in the centers of neutron stars with the masses corresponding to $n_{\rm cen}\le (2-2.5) n_0$ (the values depend on the equation of state used). Actually, $O$He already dissociates not far from the crust-core boundary (for $n\ge n_0$). Indeed, for $n\ge n_0$, the $O$He Bohr radius $a_{O{\rm He}}\approx 2$ fm becomes larger than the typical distance between the nucleons and the $O$He melts, owing to the Mott transition. $O$-particles are released from $O$He for $r<R_{\rm Mott}$. Since $GMm_O^{}/R_{\rm Mott}\gg p^2_{{\rm F},n}/(2m_\mathrm{n})\gg 16$ MeV ($R_{\rm Mott}\sim 10$ km as the neutron star radius $R$), being released, the $O$-particles dive down toward the neutron star center.

Because of the charge-asymmetric nature of $O$-particles, corresponding to the absence of their annihilation, the number of $O$-particles in a star, $N_O$, obeys the equation $dN_O/dt=C_{\rm capt}$, where $C_{\rm capt}$ is the $O$He capture rate through scattering by baryons. The capture can occur only when the momentum transfer is larger than the difference between the baryon Fermi momentum and the momentum of the rescattered baryon. For $m_O^{}\gg m_\mathrm{N}^{}$, one gets \cite{Gould:1987ju,Gould:1987ir,McDermott:2011jp}
\begin{eqnarray}\label{escape}
\frac{dC_{\rm cap}}{d\Omega_3}\simeq \sum_b \sqrt{\frac{6}{\pi}}\frac{\rho_{O{\rm He}}^{}(r)}{m_O^{}}\frac{v^2(r)}{\bar{v}^2}n_b (r)(\bar{v}\sigma_{O{\rm He},b})\xi_b\left[ 1-\frac{1-e^{-B_b^2}}{B_b^2} \right]\,,
\end{eqnarray}
where $\Omega_3$ is the neutron-star volume; $\rho_{O{\rm He}}^{}(r)$ is the ambient $O{\rm He}$ mass density; $n_b (r)$ is the number density of the baryon species $b=(n,p, H,...)$, $H= \Lambda ,\Sigma,\Xi$; $\bar{v}$ is the $O$He-velocity dispersion around the neutron star; $v(r)$ is the escape velocity of the neutron star at the given radius $r$; $\sigma_{O{\rm He},b}$ is the effective scattering cross section between $O$He and the baryon $b$ in the neutron star; $\xi_b=\mbox{min}\{\delta p_b /p_{{\rm F},b},1\}$ takes into account the neutron degeneracy effect on the capture; $\delta p_b\simeq \sqrt{2} m_{\rm red}v_{\rm esc}$; $m_{\rm red}$ is the reduced $O$He--baryon mass, $m_{\rm red}\simeq m_\mathrm{N}^{}$; $p_{{\rm F},b}^{}$ is the Fermi momentum of the $b$ baryon; and $B_b^2 \simeq 6 m_b v^2(r)/(m_O^{} \bar{v}^2)$.

Near the boundary of the neutron-star crust core, $n\sim n_0$ and $n_\mathrm{n}\gg n_\mathrm{p}$, $n_H^{}=0$. Typically \cite{McDermott:2011jp}, $v_{\rm esc}\sim v(r\sim R)\sim p_{{\rm F},n}^{}/m_\mathrm{N}^{}\sim 10^5$ km$/$s for $n\simeq n_\mathrm{n}\sim n_0$, and thus $\xi_b \sim 1$, $\bar{v}\simeq 250$ km$/$s, and thereby $B_b\gg 1$. Then, Eq.~\eqref{escape} simplifies as \begin{eqnarray}\label{escapeav}
C_{\rm cap}\sim\frac{\rho_{O{\rm He}}}{m_{O}}\frac{v_{\rm esc}^2}{\bar{v}^2}\bar{v}\sigma_{O{\rm He},n}N_n\,, \quad {\rm and}\quad N_O\simeq C_{\rm cap} t\,.
\end{eqnarray}
The maximum value for $\sigma_{O{\rm He},n}^{}$ is $\pi R^2/N_\mathrm{n}$, and we are able to estimate a maximum number of $N_O^{\rm max}$ and a maximum $O$-nuclearite mass accumulated in the center of a neutron star of the given age
$$
N_O^{\rm max}\sim 10^{39}\frac{t}{10^{10}{\rm yr}}\,,\quad M_{O{\rm-nuclearite}}^{\rm max}\sim 10^{18}\frac{t}{10^{10}{\rm yr}}{\rm g}\,.
$$
For a self-bound $O$-nuclearite, its radius is found from $n\,\Omega_{O{\rm-nuclearite}}=N_O$, and we get $R^{\rm max}\sim {\rm cm}$. Thus, a $\sim 10^7$ times enhanced $O$He flux onto the neutron star is needed, compared to that we have used in above estimates, to accumulate inside the old neutron star of the age $\sim 10^{10}$ yr, a mass $M_{O{\rm-nuclearite}}\sim M_{\odot}$.

We may perform similar estimations for the red giants, which during their evolution also  may accumulate $O$He matter in the star centers. Taking $R\sim 10^9 $ km, $M\sim 0.5 M_{\odot}$, $t_{\rm life}\sim 10^8$yr, $v_{\rm esc}\sim \bar{v}$, we estimate $N_O^{\rm max}\sim 10^{46}$ and $M_{O{\rm -nuclearite}}^{\rm max}\sim 10^{25}$ g. Similar estimates are valid for red supergiants. The $O$He nugget, being formed in the center of the star, awaits then the supernova explosion. When nucleons begin to fall to the center, the self-bounded $O$-nuclearite might be formed.

\section{Conclusion}
\label{sec:sec5}

With the lack of evidence for WIMPs in direct and indirect searches for dark matter, the fields of study of possible dark matter physics should be strongly extended. Dark atoms of $O$He are of special interest in view of the minimal involvement of new physics in their properties. The hypothesis on stable double charged particle constituents of dark atoms sheds new light on the strategy of dark matter studies, offering a nontrivial explanation for the puzzles of direct and indirect dark matter searches. In particular, in the context of this hypothesis, collider searches for dark matter are not related to the effect of missing mass, momentum, and energy, but are related to the search for stable double charged particles. Astrophysical indirect effects of $O$He dark matter are related to radiation from $O$He excitation in collisions in the center of Galaxy. It can explain the excess of the positronium annihilation line, observed by INTEGRAL in the galactic bulge, provided that the mass of the double charged $O$ particle is near 1.25 TeV, that is within the reach of search for such particle at the LHC.

However simple in the description of new physics, the old-fashioned and seemingly well-known nuclear and atomic physics turn out to be nontrivial and rather complicated in the description of dark atoms and their interaction with matter. Nuclear physics of $O$He atoms is still unclear and remains an open problem of this approach. The crucial point is the existence of a potential barrier in the interaction of $O$He with nuclei. If such a barrier exists in the $O$He interaction with sodium nuclei, the capture of the Na nucleus by $O$He to a low-energy bound state beyond nuclear radii can explain the positive effect of direct dark matter searches for the annual modulation signal in DAMA/NaI and DAMA/LIBRA experiments. Annual modulation follows in this explanation from the annual modulation of the $O$He concentration in the matter of the detector, while small recoil energy explains the absence of positive effects in other experiments. The rate of capture is determined by electric dipole transition, which is strongly suppressed in cryogenic detectors, while the absence of a low-energy bond state in $O$He interaction with heavy nuclei makes it impossible to test this hypothesis in detectors with heavy element content, like liquid xenon. On the other hand, if such barrier does not exist or is not efficient, inelastic collisions dominate in $O$He-nucleus interactions and overproduction of anomalous isotopes inevitably rules out the $O$He dark matter hypothesis.

The formation of an $O$H$^{-}$ ion in proton capture by $O^{--}$ may lead to another potential problem for the $O$He scenario. The abundance of such ions is severely constrained by searches for stable charged massive particles and anomalous isotopes in sea water \cite{Smith.NPB.1979,Smith.NPB.1982,Hemmick.PRD.1990,Verkerk.PRL.1992,Yamagata.PRD.1993,Kudo.PLB.2001}. Production of such ions in the early Universe is strongly suppressed, since all the free $O^{--}$ are captured by primordial helium before proton capture becomes possible. However, in the Galaxy, $O$He destructions in stars and in cosmic rays can release free $O^{--}$, which can be captured by protons, forming $O$H$^{-}$ ions and an anomalous $-2$ charged component of cosmic rays. In principle, the capture of such components by Earth can lead to a dangerous amount of anomalous isotopes in sea water, but the corresponding analysis, involving detailed study of $O$He evolution in the Galaxy, goes beyond the scope of the present work.

Putting aside these problems, we turn here to the extension of studies of possible effects of $O$He in nuclear matter and astrophysical conditions. We proposed the possibility of the existence of stable $O$-nuclearites and discussed various mechanisms for their formation. Observation of $O$-nuclearites, in which dark matter is bound with the normal nuclear matter, would be an important event that could provide us additional information on the possibility of the existence of dark $O$He atoms of dark matter and their properties.

\section*{Acknowledgments}

The work of V.A.G.\ on studies of $O$-nuclearites was supported by the MEPhI Academic Excellence Project under Contract No.\ 02.a03.21.0005. The work of D.N.V.\ on studies of the charged massive particle binding in $O$-nuclearites was also supported by the Ministry of Education and Science of the Russian Federation within the state assignment, Project No~3.6062.2017/6.7. The work of M.Yu.K.\ on studies of effects of $O$He dark matter was supported by grant of the Russian Science Foundation (Project No-18-12-00213).

\vspace{10mm}

\hrule

\vspace{20mm}

\begin{figure}[h!]
\centering
\includegraphics[width=0.2\textwidth]{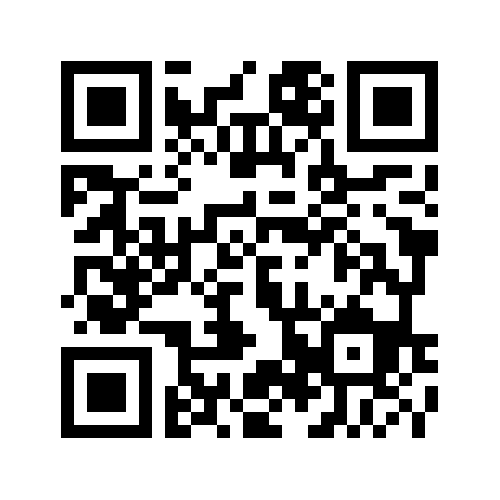}
\end{figure}


\begin{thebibliography}{99}

\bibitem{Ade:2015xua}
P.~A.~R.~Ade {\it et al.} (Planck Collaboration), {Planck 2015 results. XIII. Cosmological parameters}, \href{https://doi.org/10.1051/0004-6361/201525830}{{Astron.\ Astrophys.} {\bf 594}, A13 (2016)} [\href{https://arxiv.org/abs/1502.01589}{\tt arXiv:1502.01589}].

\bibitem{Arcadi.2018}
G.~Arcadi {\it et al.}, {The waning of the WIMP? A review of models, searches, and constraints}, \href{https://doi.org/10.1140/epjc/s10052-018-5662-y}{Eur.\ Phys.\ J.\ C {\bf 78}, 203 (2018)} [\href{https://arxiv.org/abs/1703.07364}{\tt arXiv:1703.07364}].

\bibitem{Belotsky.2014}
K.~M.~Belotsky {\it et al.}, {Signatures of primordial black hole dark matter}, \href{https://doi.org/10.1142/S0217732314400057}{Mod.\ Phys.\ Lett.\ A {\bf 29}, 1440005 (2014)} [\href{https://arxiv.org/abs/1410.0203}{\tt arXiv:1410.0203}].

\bibitem{Carr.2016}
B.~Carr, F.~K\"uhnel, and M.~Sandstad, {Primordial black holes as dark matter}, \href{https://doi.org/10.1103/PhysRevD.94.083504}{Phys.\ Rev.\ D {\bf 94}, 083504 (2016)} [\href{https://arxiv.org/abs/1607.06077}{\tt arXiv:1607.06077}].

\bibitem{Carr.PRD.2010}
B.~J.~Carr, K.~Kohri, Yu.~Sendouda, and J.~Yokoyama, {New cosmological constraints on primordial black holes}, \href{https://doi.org/10.1103/PhysRevD.81.104019}{Phys.\ Rev.\ D {\bf 81}, 104019 (2010)} [\href{https://arxiv.org/abs/0912.5297}{\tt arXiv:0912.5297}].

\bibitem{DMRev}
M.~Yu.~Khlopov, {Fundamental particle structure in the cosmological dark matter}, \href{https://doi.org/10.1142/S0217751X13300421}{Int.\ J.\ Mod.\ Phys.\ A {\bf 28}, 1330042 (2013)} [\href{https://arxiv.org/abs/1311.2468}{\tt arXiv:1311.2468}].

\bibitem{Aprile:2009zzd}
E.~Aprile and S.~Profumo, {Focus on dark matter and particle physics}, \href{https://doi.org/10.1088/1367-2630/11/10/105002}{New J.\ Phys.\ {\bf 11}, 105002 (2009)}.

\bibitem{Feng:2010gw}
J.~L.~Feng, {Dark matter candidates from particle physics and methods of detection}, \href{https://doi.org/10.1146/annurev-astro-082708-101659}{Annu.\ Rev.\ Astron.\ Astrophys.\ {\bf 48}, 495 (2010)} [\href{https://arxiv.org/abs/1003.0904}{\tt arXiv:1003.0904}].

\bibitem{Glashow:2005jy}
S.~L.~Glashow, {A sinister extension of the Standard Model to $SU(3) \times SU(2) \times SU(2) \times U(1)$}, \href{https://arxiv.org/abs/hep-ph/0504287}{\tt hep-ph/0504287}.

\bibitem{BKSR1}
D.~Fargion and M.~Yu.~Khlopov, {Tera-leptons' shadows over Sinister Universe}, \href{https://doi.org/10.1134/S0202289313040063}{{Gravitation Cosmol.} {\bf 19}, 219 (2013)} [\href{https://arxiv.org/abs/hep-ph/0507087}{\tt hep-ph/0507087}].

\bibitem{Khlopov:2005ew}
M.~Yu.~Khlopov, {\it Composite dark matter from 4th generation}, Pis'ma Zh.\ Eksp.\ Teor.\ Fiz.\ {\bf 83}, 3 (2006) [\href{https://doi.org/10.1134/S0021364006010012}{{JETP Lett.} {\bf 83}, 1 (2006)}] [\href{https://arxiv.org/abs/astro-ph/0511796}{\tt astro-ph/0511796}].

\bibitem{Khlopov:2010ik}
M.~Yu.~Khlopov, A.~G.~Mayorov, and E.~Yu.~Soldatov, {The dark atoms of dark matter}, Prespacetime J.\ {\bf 1}, 1403 (2010) [\href{https://arxiv.org/abs/1012.0934}{\tt arXiv:1012.0934}].

\bibitem{Khlopov:2011me}
M.~Yu.~Khlopov, A.~G.~Mayorov, and E.~Yu.~Soldatov, {Towards nuclear physics of OHe dark matter}, Bled Workshops Phys.~{\bf 12}, 94 (2011) [\href{https://arxiv.org/abs/1111.3577}{\tt arXiv:1111.3577}].

\bibitem{gKhlopov}
M.~Yu.~Khlopov, {Physics of dark matter in the light of dark atoms}, \href{https://doi.org/10.1142/S0217732311037194}{{Mod.\ Phys.\ Lett.\ A} {\bf 26}, 2823 (2011)} [\href{https://arxiv.org/abs/1111.2838}{\tt arXiv:1111.2838}].

\bibitem{khlopov.proc.2014}
J.-R.~Cudell, M.~Khlopov, and Q.~Wallemacq, {Some potential problems of OHe composite dark matter}, Bled Workshops Phys.\ {\bf 15}, 66 (2014) [\href{https://arxiv.org/abs/1412.6030}{\tt arXiv:1412.6030}].

\bibitem{khlopov.proc.2015}
M.~Yu.~Khlopov, {10 years of dark atoms of composite dark matter}, Bled Workshops Phys.\ {\bf 16}, 71 (2015) [\href{https://arxiv.org/abs/1512.01081}{\tt arXiv:1512.01081}].

\bibitem{khlopov.ijmpd.2015}
J.~R.~Cudell and M.~Khlopov, {Dark atoms with nuclear shell: A status review}, \href{https://doi.org/10.1142/S0218271815450078}{Int.\ J.\ Mod.\ Phys.\ D {\bf 24}, 1545007 (2015)}.

\bibitem{Wallemacq:2015cjr}
Q.~Wallemacq, {Composite dark matter and direct-search experiments}, \href{https://doi.org/10.1142/S0218271815450066}{Int.\ J.\ Mod.\ Phys.\ D {\bf 24}, 1545006 (2015)} [\href{https://arxiv.org/abs/1512.05898}{\tt arXiv:1512.05898}].

\bibitem{probes}
M.~Yu.~Khlopov, {Probes for dark matter physics}, \href{https://doi.org/10.1142/S0218271818410134}{Int.\ J.\ Mod.\ Phys.\ D {\bf 27}, 1841013 (2018)} [\href{https://arxiv.org/abs/1802.10184}{\tt arXiv:1802.10184}].

\bibitem{KK}
M.~Yu.~Khlopov and C.~Kouvaris, {Strong interactive massive particles from a strong coupled theory}, \href{https://doi.org/10.1103/PhysRevD.77.065002}{Phys.\ Rev.\ D {\bf 77}, 065002 (2008)} [\href{https://arxiv.org/abs/0710.2189}{\tt arXiv:0710.2189}].

\bibitem{KK1}
M.~Yu.~Khlopov and C.~Kouvaris, {Composite dark matter from a model with composite Higgs boson}, \href{https://doi.org/10.1103/PhysRevD.78.065040}{Phys.\ Rev.\ D {\bf 78}, 065040 (2008)} [\href{https://arxiv.org/abs/0806.1191}{\tt arXiv:0806.1191}].

\bibitem{CMB}
K.~K.~Boddy, V.~Gluscevic, V.~Poulin, E.~D.~Kovetz, M.~Kamionkowski, and R.~Barkana, {A critical assessment of CMB limits on dark matter-baryon scattering: New treatment of the relative bulk velocity},  \href{https://doi.org/10.1103/PhysRevD.98.123506}{Phys.\ Rev.\ D {\bf 98}, 123506 (2018)} [\href{https://arxiv.org/abs/1808.00001}{\tt arXiv:1808.00001}].

\bibitem{FS}
S.~Hirano and V.~Bromm, {Baryon-dark matter scattering and first star formation}, \href{https://doi.org/10.1093/mnrasl/sly132}{Mon.\ Not.\ R.\ Astron.\ Soc.\ {\bf 480}, L85 (2018)} [\href{https://arxiv.org/abs/1803.10671}{\tt arXiv:1803.10671}].

\bibitem{BC}
A.~Robertson, R.~Massey, and V.~Eke, {What does the Bullet Cluster tell us about self-interacting dark matter?}, \href{https://doi.org/10.1093/mnras/stw2670}{Mon.\ Not.\ R.\ Astron.\ Soc.\ {\bf 465}, 569 (2017)} [\href{https://arxiv.org/abs/1605.04307}{\tt arXiv:1605.04307}].

\bibitem{Migdal:1971cu}
A.~B.~Migdal, {Stability of vacuum and limiting fields}, \href{http://inspirehep.net/record/72370}{Zh.\ Eksp.\ Teor.\ Fiz.\ {\bf 61}, 2209 (1971)} [\href{http://www.jetp.ac.ru/cgi-bin/index/r/61/6/p2209?a=list}{Sov.\ Phys.\ JETP {\bf 34}, 1184 (1972)}].

\bibitem{Migdal:1974yx}
A.~B.~Migdal, {Meson condensation and anomalous nuclei}, \href{https://doi.org/10.1016/0370-2693(74)90081-1}{Phys.\ Lett.\ {\bf 52B}, 172 (1974)}.

\bibitem{Migdal:1978az}
A.~B.~Migdal, {Pion fields in nuclear matter}, \href{https://doi.org/10.1103/RevModPhys.50.107}{Rev.\ Mod.\ Phys.\ {\bf 50}, 107 (1978)}.

\bibitem{Migdal:1990vm}
A.~B.~Migdal, E.~E.~Saperstein, M.~A.~Troitsky, and D.~N.~Voskresensky, {Pion degrees of freedom in nuclear matter}, \href{https://doi.org/10.1016/0370-1573(90)90132-L}{Phys.\ Rep.\ {\bf 192}, 179 (1990)}.

\bibitem{Lee:1974ma}
T.~D.~Lee and G.~C.~Wick, {Vacuum stability and vacuum excitation in a spin-0 field theory}, \href{https://doi.org/10.1103/PhysRevD.9.2291}{Phys.\ Rev.\ D {\bf 9}, 2291 (1974)}.

\bibitem{Lee:1974kn}
T.~D.~Lee, {Abnormal nuclear states and vacuum excitation}, \href{https://doi.org/10.1103/RevModPhys.47.267}{Rev.\ Mod.\ Phys.\ {\bf 47}, 267 (1975)}.

\bibitem{Bodmer:1971we}
A.~R.~Bodmer, {Collapsed nuclei}, \href{https://doi.org/10.1103/PhysRevD.4.1601}{Phys.\ Rev.\ D {\bf 4}, 1601 (1971)}.

\bibitem{Migdal:1977rn}
A.~B.~Migdal, V.~S.~Popov, and D.~N.~Voskresensky, {Distribution of vacuum charge near supercharged nuclei}, Zh.~Eksp.~Teor.~Fiz.~{\bf 72}, 834 (1977) [Sov.~Phys.~JETP {\bf 45}, 436 (1977)].

\bibitem{Voskresensky:1977mz}
D.~N.~Voskresensky, G.~A.~Sorokin, and A.~I.~Chernoutsan, {Charge distribution in anomalous nuclei}, Pis'ma Zh.\ Eksp.\ Teor.\ Fiz.\ {\bf 25}, 495 (1977) [JETP Lett.\ {\bf 25}, 465 (1977)].

\bibitem{Voskresensky:1978uf}
D.~N.~Voskresensky and A.~I.~Chernoutsan, {Condensation of pions in electric field of supercharged nucleus}, Yad.\ Fiz.\ {\bf 27}, 1411 (1978) [Sov.\ J.\ Nucl.\ Phys.\ {\bf 27}, 742 (1978)].

\bibitem{Voskresensky1977}
D.~N.~Voskresensky, {Charge distribution in anomalous nuclei}, Ph.D.\ thesis, MEPhI, Moscow, 1977.

\bibitem{Kolomeitsev:2002gd} 
E.~E.~Kolomeitsev and D.~N.~Voskresensky, {Resonance states below the pion-nucleon threshold and their consequences for nuclear systems}, \href{https://doi.org/10.1103/PhysRevC.67.015805}{Phys.\ Rev.\ C {\bf 67}, 015805 (2003)} [\href{https://arxiv.org/abs/nucl-th/0207091}{\tt nucl-th/0207091}].

\bibitem{Witten:1984rs}
E.~Witten, {Cosmic Separation of Phases}, \href{https://doi.org/10.1103/PhysRevD.30.272}{Phys.\ Rev.\ D {\bf 30}, 272 (1984)}.

\bibitem{Alcock:1986hz}
C.~Alcock, E.~Farhi, and A.~Olinto, {Strange stars}, \href{https://doi.org/10.1086/164679}{{Astrophys.\ J.} {\bf 310}, 261 (1986)}.

\bibitem{DeRujula:1984axn}
A.~De Rujula and S.~L.~Glashow, {Nuclearites --- A novel form of cosmic radiation}, \href{https://doi.org/10.1038/312734a0}{{Nature (London)} {\bf 312}, 734 (1984)}.

\bibitem{DeRujula:1989fe} A.~De Rujula, S.~L.~Glashow, and U.~Sarid,  {Charged dark matter}, \href{https://doi.org/10.1016/0550-3213(90)90227-5}{Nucl.\ Phys.\ {\bf B333}, 173 (1990)}.

\bibitem{Weber}
F.~Weber, {\it Pulsars as
Astrophysical Laboratories for Nuclear and Particle Physics} (Institute of Physics Publishing, Bristol, 1999).

\bibitem{Ivanov:2005be}
Yu.~B.~Ivanov, A.~S.~Khvorostukhin, E.~E.~Kolomeitsev, V.~V.~Skokov, V.~D.~Toneev, and D.~N.~Voskresensky, {Lattice QCD constraints on hybrid and quark stars}, \href{https://doi.org/10.1103/PhysRevC.72.025804}{Phys.\ Rev.\ C {\bf 72}, 025804 (2005)} [\href{https://arxiv.org/abs/astro-ph/0501254}{\tt astro-ph/0501254}].

\bibitem{Boeckel:2010hm}
T.~Boeckel, M.~Hempel, I.~Sagert, G.~Pagliara, B.~Sa'd, and J.~Schaffner-Bielich, {Strangeness in astrophysics and cosmology}, \href{https://doi.org/10.1088/0954-3899/37/9/094005}{J.\ Phys.\ G {\bf 37}, 094005 (2010)} [\href{https://arxiv.org/abs/1002.1793}{\tt arXiv:1002.1793}].

\bibitem{Dondi:2016yjl}
N.~A.~Dondi, A.~Drago, and G.~Pagliara, {Conditions for the existence of stable strange quark matter}, \href{https://doi.org/10.1051/epjconf/201713709004}{EPJ Web Conf.\ {\bf 137}, 09004 (2017)} [\href{https://arxiv.org/abs/1612.00755}{\tt arXiv:1612.00755}].

\bibitem{Kouvaris.PRD.2014}
C.~Kouvaris and P.~Tinyakov, {Growth of black holes in the interior of rotating neutron stars}, \href{https://doi.org/10.1103/PhysRevD.90.043512}{Phys.\ Rev.\ D {\bf 90}, 043512 (2014)} [\href{https://arxiv.org/abs/1312.3764}{\tt arXiv:1312.3764}]. 

\bibitem{Cahn}
R.~N.~Cahn and S.~L.~Glashow, {Chemical signatures for superheavy elementary Particles}, \href{https://doi.org/10.1126/science.213.4508.607}{Science {\bf 213}, 607 (1981)}.

\bibitem{Pospelov}
M.~Pospelov, {Particle Physics Catalysis of Thermal Big Bang Nucleosynthesis}, \href{https://doi.org/10.1103/PhysRevLett.98.231301}{Phys.\ Rev.\ Lett.\ {\bf 98}, 231301 (2007)} [\href{https://arxiv.org/abs/hep-ph/0605215}{\tt hep-ph/0605215}].

\bibitem{Kohri}
K.~Kohri and F.~Takayama, {Big bang nucleosynthesis with long-lived charged massive particles}, \href{https://doi.org/10.1103/PhysRevD.76.063507}{Phys.\ Rev.\ D {\bf 76}, 063507 (2007)} [\href{https://arxiv.org/abs/hep-ph/0605243}{\tt hep-ph/0605243}].

\bibitem{Klahn:2006ir}
T.~Klahn {\it et al.}, {Constraints on the high-density nuclear equation of state from the phenomenology of compact stars and heavy-ion collisions}, \href{https://doi.org/10.1103/PhysRevC.74.035802}{Phys.\ Rev.\ C {\bf 74}, 035802 (2006)} [\href{https://arxiv.org/abs/nucl-th/0602038}{\tt nucl-th/0602038}].

\bibitem{Chin:1979yb}
S.~A.~Chin and A.~K.~Kerman, {Possible Long-Lived Hyperstrange Multiquark Droplets}, \href{https://doi.org/10.1103/PhysRevLett.43.1292}{Phys.\ Rev.\ Lett.\ {\bf 43}, 1292 (1979)}.

\bibitem{Gould:1987ju}
A.~Gould, {WIMP distribution in and evaporation from the Sun}, \href{https://doi.org/10.1086/165652}{Astrophys.\ J.\ {\bf 321}, 560 (1987)}.

\bibitem{Gould:1987ir}
A.~Gould, {Resonant enhancements in WIMP capture by the Earth}, \href{https://doi.org/10.1086/165653}{Astrophys.\ J.\ {\bf 321}, 571 (1987)}.

\bibitem{McDermott:2011jp}
S.~D.~McDermott, H.~B.~Yu, and K.~M.~Zurek, {Constraints on scalar asymmetric dark matter from black hole formation in neutron stars}, \href{https://doi.org/10.1103/PhysRevD.85.023519}{Phys.\ Rev.\ D {\bf 85}, 023519 (2012)} [\href{https://arxiv.org/abs/1103.5472}{\tt arXiv:1103.5472}].

\bibitem{Smith.NPB.1979}
P.~F.~Smith and J.~R.~J.~Bennett, {A search for heavy stable particles}, \href{https://doi.org/10.1016/0550-3213(79)90006-3}{Nucl.\ Phys.\ {\bf B149}, 525 (1979)}.

\bibitem{Smith.NPB.1982}
P.~F.~Smith {\it et al.}, {A search for anomalous hydrogen in enriched $D_2O$, using a time-of-flight spectrometer}, \href{https://doi.org/10.1016/0550-3213(82)90271-1}{Nucl.\ Phys.\ {\bf B206}, 333 (1982)}.

\bibitem{Hemmick.PRD.1990}
T.~K.~Hemmick {\it et al.}, {Search for low-Z nuclei containing massive stable particles}, \href{https://doi.org/10.1103/PhysRevD.41.2074}{Phys.\ Rev.\ D {\bf 41}, 2074 (1990)}.

\bibitem{Verkerk.PRL.1992}
P.~Verkerk {\it et al.}, {Search for superheavy hydrogen in sea water}, \href{https://doi.org/10.1103/PhysRevLett.68.1116}{Phys.\ Rev.\ Lett.\ {\bf 68}, 1116 (1992)}.

\bibitem{Yamagata.PRD.1993}
T.~Yamagata, Y.~Takamori, and H.~Utsunomiya, {Search for anomalously heavy hydrogen in deep sea water at 4000 m}, \href{https://doi.org/10.1103/PhysRevD.47.1231}{Phys.\ Rev.\ D {\bf 47}, 1231 (1993)}.

\bibitem{Kudo.PLB.2001}
A.~Kudo and M.~Yamaguchi, {Inflation with low reheat temperature and cosmological constraint on stable charged massive particles}, \href{https://doi.org/10.1016/S0370-2693(01)00938-8}{Phys.\ Lett.\ B {\bf 516}, 151 (2001)} [\href{https://arxiv.org/abs/hep-ph/0103272}{\tt hep-ph/0103272}].

\end{thebibliography}
\end{document}